\documentclass[useASM]{mn2e}

\usepackage{graphicx}

\begin{document}

\title[The remnants of binary neutron star merger]
{Diagnosing the remnants of binary neutron star merger from GW170817/GRB170817A event}

\author[L\"{u} et al.]
{Hou-Jun L\"{u}$^{1}$\thanks{E-mail: lhj@gxu.edu.cn}, Jun Shen$^{1}$, Lin Lan$^{1}$,
Jared Rice$^{2}$, Wei-Hua Lei$^{3}$, and En-Wei Liang$^{1}$\\
 $^1$Guangxi Key Laboratory for Relativistic Astrophysics, School of Physical Science and
 Technology, Guangxi University, Nanning 530004, China\\
 $^2$Department of Physics, Texas State University, San Marcos, TX 78666, USA\\
 $^3$School of Physics, Huazhong University of Science and Technology, Wuhan 430074,
China\\}
\maketitle

\label{firstpage}
\begin{abstract}
The event GW170817/GRB 170817A, discovered via the successful joint observation of its
gravitational wave radiation and its multi-wavelength electromagnetic counterparts, was the first
definite ``smoking-gun'' from the merger of two neutron stars (NSs). However, the remnant of the
merger remains unknown. Piro et al. recently claimed that a low-significance X-ray variability in
GRB 170817A. By systematically comparing the properties of variability in the afterglow of GRB
170817A and X-ray flares in GRB afterglows, we find that this X-ray variability seems to share
similar statistical correlations with X-ray flares in GRB afterglows. We further investigate
several possible merger product scenarios to see whether they can produce the observed X-ray
variability in GRB 170817A. The first scenario invokes a stable magnetar as the central engine
producing the later X-ray variability via differential rotation or fall-back accretion onto the NS.
The second scenario invokes a black hole as the central engine with a fall-back accretion process.
The final scenario is a central engine with a long-lived supra-massive NS. We find that the first
two scenarios have difficulty producing the later X-ray variability, which requires either an
impractical NS magnetic field or an extraordinarily large stellar envelope and an extremely long
accretion timescale. However, the third scenario seems to be consistent with observations, and the
later X-ray variability can be produced by the magnetosphere which is expelled following the
collapse of the NS with a $B_p\in(3.6, 13.5)\times10^{13}$ G.
\end{abstract}
\begin{keywords}
gravitational waves-gamma-ray burst: general-neutron stars
\end{keywords}

\section {Introduction}
On August 17, 2017, the first direct detection of gravitational waves (GW170817) originating from
the merger of a binary neutron star (NS) system was achieved by Advanced LIGO and Advanced VIRGO
(Abbott et al. 2017a,b; Goldstein et al. 2017; Savchenko et al. 2017). More interestingly, a short
gamma-ray burst (GRB170817A) 1.7 s after the binary's coalescence time and the confirmed kilonova
(AT2017gfo) 11 hours after the merger, were discovered by {\em Fermi}, INTEGRAL, and other optical
telescopes with an inferred sky location consistent with that measured for GW170817 (Coulter et al.
2017; Pian et al. 2017; Covino et al. 2017; Kasen et al. 2017; Zhang et al. 2018).

The outcome of a double NS merger depends on the total mass of the system and the poorly known NS
equation of state (EOS; Lasky et al. 2014; Li et al. 2016). From the theoretical point of view,
there are four different types of merger remnants that are discussed in the literature (Rosswog et
al. 2000; Dai et al. 2006; Fan \& Xu 2006; Gao \& Fan 2006; Rezzolla et al. 2010; Giacomazzo \&
Perna 2013; Zhang 2013; Lasky et al. 2014; Rosswog et al. 2014; Radice et al. 2018). The first type
is a prompt black hole (BH) which forms if the mass of the binary is larger than 1.7 times the
maximum mass for a non-rotating NS (Hotokezaka et al. 2011; Bauswein et al. 2013). The second is a
hyper-massive NS with a rest mass that exceeds the maximum allowable for a uniformly rotating NS,
which can then survive for $\sim100$ ms before collapsing into a BH (Baumgarte et al. 2000; Shibata
\& Taniguchi 2006; Palenzuela et al. 2015). The third is a supra-massive NS supported by rigid
rotation which survives to much longer times (e.g., seconds to hours). The supra-massive NS is
stable if its rest mass exceeds the maximum mass of a non-rotating NS, but is lower than the
maximum mass of a rotating NS (Hotokezaka et al. 2013; L\"{u} et al. 2015; Gao et al. 2016; Foucart
et al. 2016; Kiuchi et al. 2018). The last one is stable NS when the rest mass of NS is less than
the maximum mass of a non-rotating NS (Giacomazzo \& Perna 2013).

The multi-wavelength nature of the electromagnetic (EM) counterparts to GW 170817 event is of great
interest and is playing a crucial role in constraining the progenitors or remnants of NS-NS
mergers. Especially, the kilonova AT2017gfo observed post-merger is providing an unprecedented
opportunity to probe the ingredients of the merger ejecta and the energy sources of the emission
(Kasen et al. 2017). Yu et al. (2018) proposed that the later emission of AT2017gfo is primarily
caused by delayed energy injection from a long-lived remnant NS. On the contrary, Pooley et al.
(2018) suggested that the remnant of GW170817 is most likely a black hole by comparing the measured
X-ray flux observed in the Chandra X-Ray Observatory's Director's Discretionary Time in the first
four months post-merger with that predicted for a neutron star remnant. In any case, the
post-merger product of GW 170817/GRB 170817A is still under debate based on current observational
data (Margalit \& Metzger 2017; Shibata et al. 2017; Yu et al. 2018; Pooley et al 2018; Ai et al.
2018; Geng et al. 2018; Li et al. 2018).

Recently, Piro et al. (2019) claimed that X-ray variability (analogous to X-ray flares observed in
GRB afterglows) with a low-significance temporal feature ($\sim 3.3\sigma$ confidence) was detected
by Chandra around 160 days post-merger for GW170817/GRB170817A. The similarity of the flare in
GW170817 with X-ray flares in GRBs was initially noted by Piro et al. (2019). They proposed that
the X-ray variability originated from an abrupt outflow injection from a long-lived magnetized NS
characterized by a strong toroidal component. If indeed the X-ray variability is an intrinsic
signal, it provides clues and a unique opportunity to study its origin, remnants of such binary
neutron star mergers, and implications for GRB physics. In this paper, we compare the properties of
late time X-ray variability in GRB 170817A with other X-ray flares observed in GRB afterglow in the
{\em Swift} era, and investigate the possible remnants of such binary NS merger systems and the
radiation mechanism of X-ray flares. Throughout the paper, a concordance cosmology with parameters
$H_0=71~\rm km~s^{-1}~Mpc^{-1}$, $\Omega_M=0.30$, and $\Omega_{\Lambda}=0.70$ is adopted.

\section{The properties of X-ray variability of GRB 170817A}
X-ray flares are discovered in a good fraction of both long and short GRBs in the {\em Swift} era.
Their light curves typically show a very rapid rise and fall with steep rising and decaying
indices. This suggests that the X-ray flares are likely produced with a mechanism similar to the
prompt emission, likely due to internal dissipation of long-lasting central engine activities
(Burrows et al. 2005; Ioka et al. 2005; Fan \& Wei 2005; Zhang et al. 2006; Margutti et al. 2010).
Yi et al. (2016) performed a systematic study of X-ray flares observed by {\em Swift}, and found
that several observational parameters of flares presented some empirical correlations (e.g.,
$L_{\rm p}-T_{\rm p}$). This indicated that those observed flares in different GRBs share a similar
mechanism.

It is important to determine whether or not the X-ray variability in GRB 170817A can be identified as
an X-ray flare. In order to investigate this issue, we compare the properties of X-ray variability in
GRB 170817A with X-ray flares observed in {\em Swift} era. Actually, we should first consider the short
GRBs samples to make the comparisons. However, there are limited short GRBs cases observed with
measured redshifts that contain X-ray flares in the afterglow and so it is difficult to perform a
statistical analysis. As such, we must choose X-ray flares in long GRBs to do the
comparisons\footnote{Here, we assume that the X-ray flares in both long and short GRBs share a similar
physical origin, e.g., the central engine reactivity.}.

\begin{figure*}
\includegraphics[angle=0,scale=0.8]{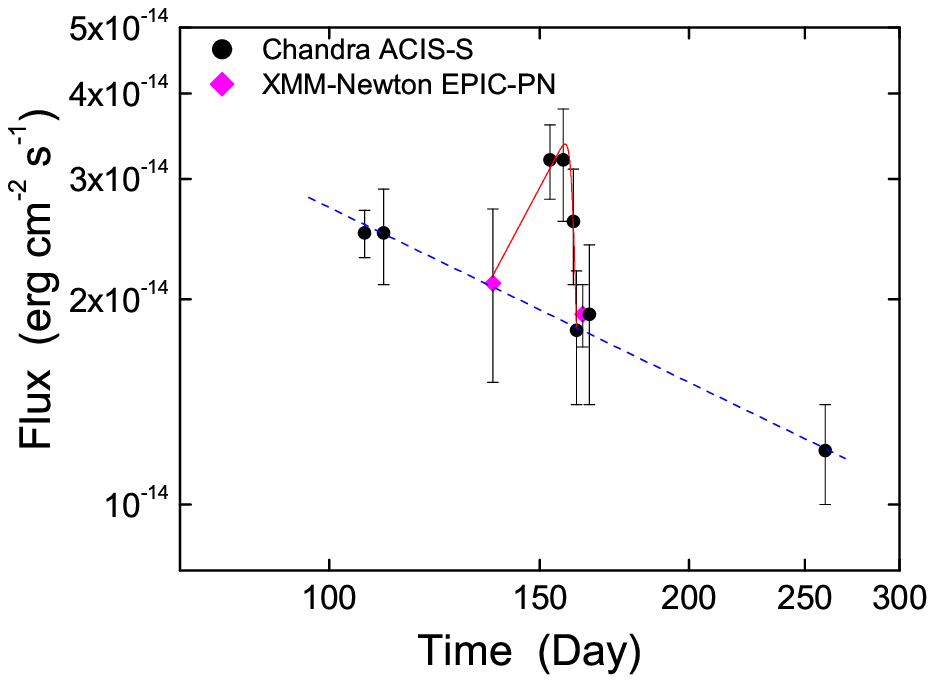}
\hfill\center \caption{X-ray afterglow of GRB 170817A after 100 days, the data are taken from Piro et al. 2019. The solid red and dashed lines show the
smooth broken power-law and power-law models fit, respectively.}
\label{fig:Xray}
\end{figure*}

We fit the X-ray variability in GRB 170817A with a smooth broken power-law function (Liang et al.
2007; Yi et al. 2016),
\begin{eqnarray}
F_1(t)=F_{01}\left [
\left (\frac{t}{t_p}\right)^{\omega\alpha_1}+\left(
\frac{t}{t_p}\right)^{\omega\alpha_2}\right]^{-1/\omega},
\label{SBPL}
\end{eqnarray}
where $\omega$ represents the sharpness of the peak and $t_p$, $\alpha_1$, and $\alpha_2$ are the
peak time, and the rising and decay slopes of X-ray variability, respectively. One has the start
time ($T_{\rm s}\sim 137$ days), end time ($T_{\rm e}\sim 161$ days), $t_p=(159.6\pm0.53)$ days,
$\alpha_1=(-3.29\pm0.89)$, and $\alpha_2=(74.3\pm31.4)$ of X-ray variability, respectively. The
fitting results are shown in Figure \ref{fig:Xray}. On the other hand, we also obtain the energy
and luminosity as well as the temporal properties of the X-ray variability, i.e., isotropic energy
$E_{\rm X, iso}=(3.13\pm1.36)\times10^{45}\rm ~erg$, isotropic luminosity $L_{\rm X,
iso}=(1.54\pm0.88)\times 10^{39}\rm ~erg~s^{-1}$, peak luminosity $L_{\rm p}= (3.19\pm1.12)\times
10^{39}\rm ~erg~s^{-1}$, duration of X-ray variability $T_{\rm d}=T_{\rm e}-T_{\rm s}\sim 24$ days,
and waiting time $T_{\rm w}\sim T_{\rm s}=137$ days\footnote{The waiting time is defined as $T_{\rm
w}= T_{\rm s,i+1}-T_{\rm s,i}$, where $T_{\rm s,i+1}$ and $T_{\rm s,i}$ are the observed start time
of the $(i+1)$th and $i$th flares, respectively (Wang \& Dai 2013; Yi et al. 2016). $i=0$ in GRB
170817A, means that the prompt emission of GRB 170817A can be adopted as the first flare.}.

\begin{figure*}
\includegraphics[angle=0,scale=0.6]{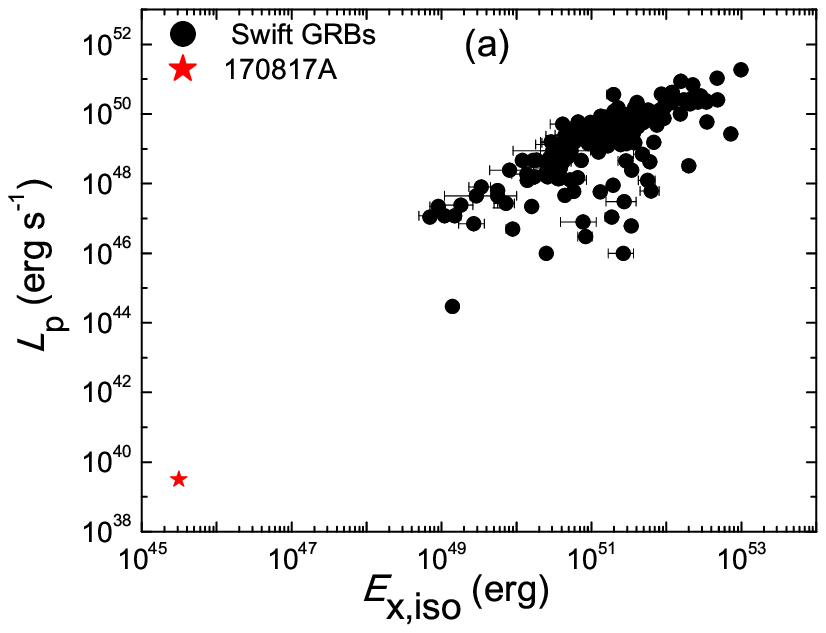}
\includegraphics[angle=0,scale=0.6]{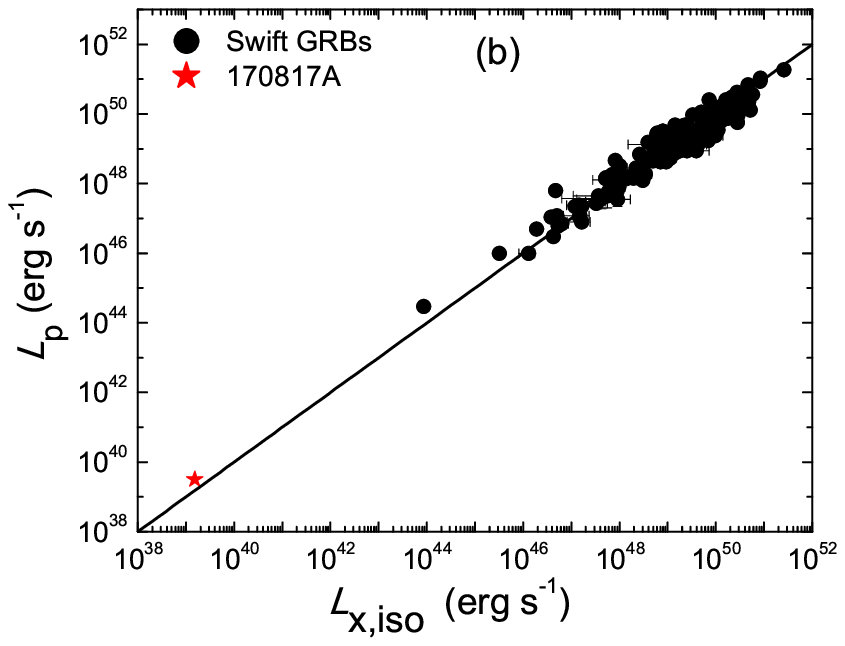}
\includegraphics[angle=0,scale=0.6]{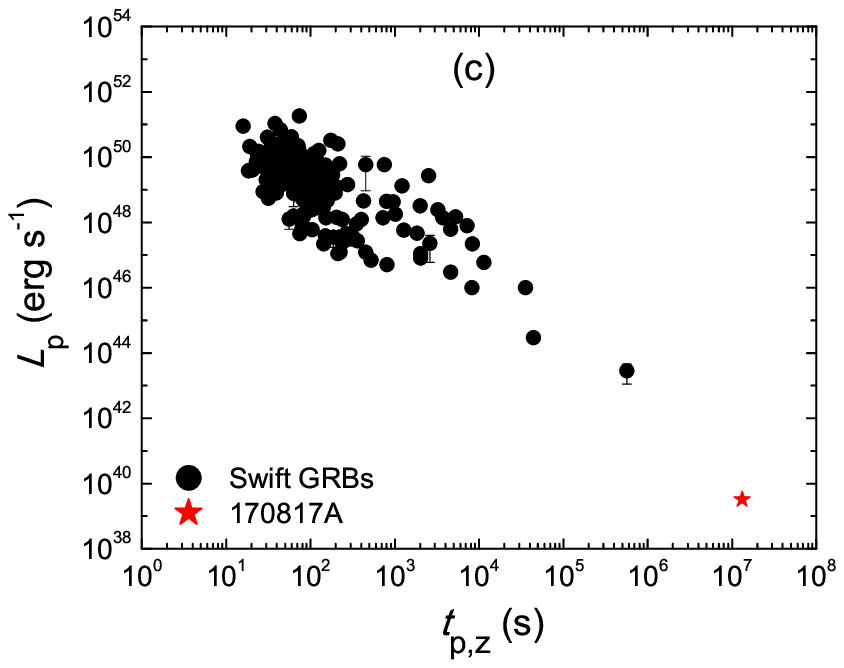}
\includegraphics[angle=0,scale=0.6]{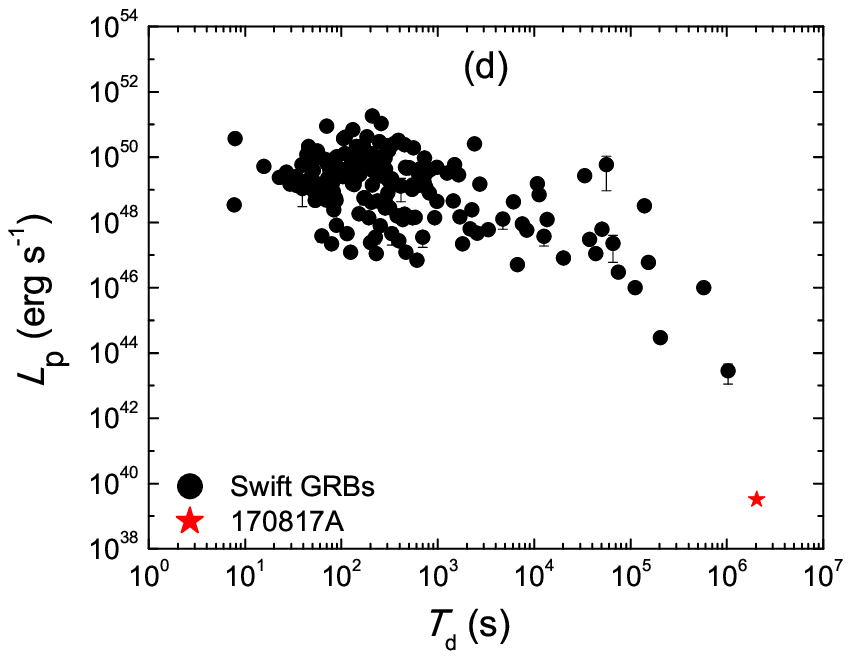}
\includegraphics[angle=0,scale=0.6]{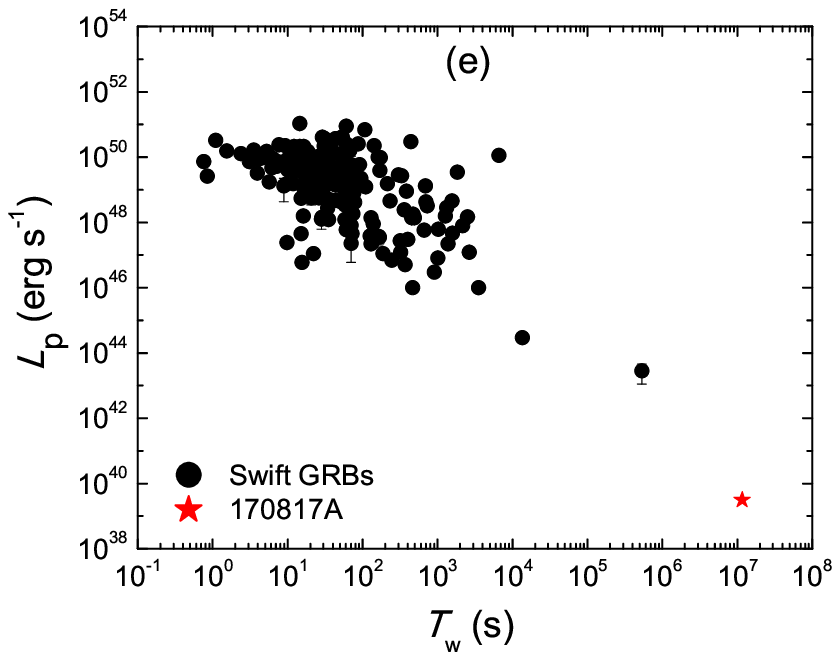}
\includegraphics[angle=0,scale=0.6]{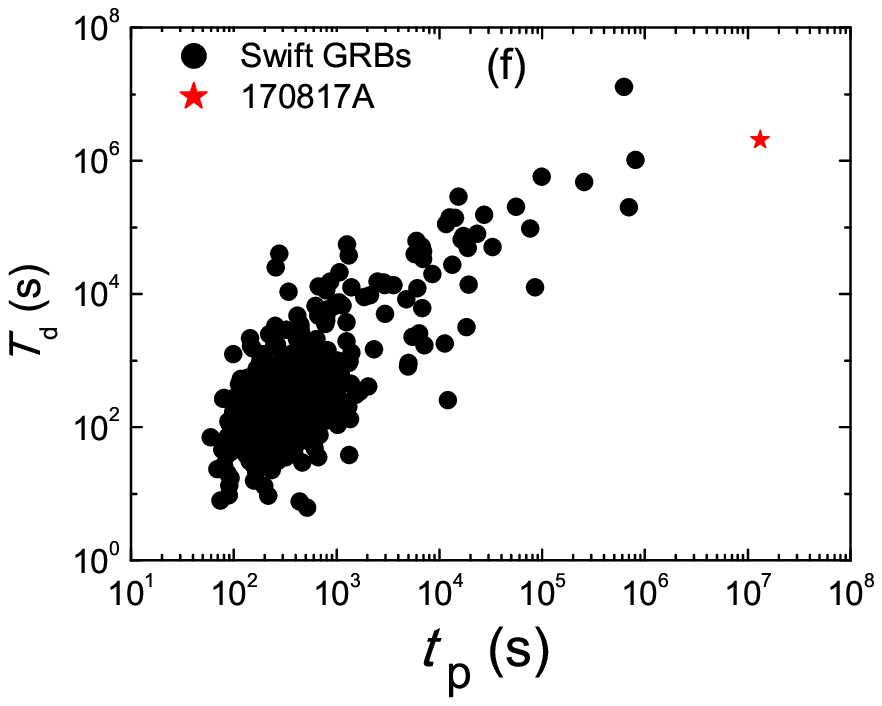}
\hfill\center
\caption{Comparisons of the peak luminosity ($L_{\rm p}$) as a function of $E_{\rm X, iso}$, $L_{\rm
X,iso}$, $t_{\rm p}$, $T_{\rm d}$, and $T_{\rm w}$ for X-ray flares in {\em Swift} GRBs (black solid
dots; taken from Yi et al. 2016) with X-ray variability in GRB 170817A (red solid star), as well
as $t_{\rm d}-T_{\rm p}$ diagram.}
\label{fig:correlation}
\end{figure*}

In Figure \ref{fig:correlation}, we make some comparisons between X-ray flares in {\em Swift} GRBs and
X-ray variability in GRB 170817A. Figure \ref{fig:correlation}(a-e) shows the peak luminosity ($L_{\rm
p}$) as function of $E_{\rm X, iso}$, $t_{\rm p,z}$, $T_{\rm d}$, and $T_{\rm w}$. The X-ray
variability of GRB 170817A seems to be consistent with those correlations with a larger scatter.
Especially, it follows a well-known $L_{\rm p}-t_{\rm p}$ correlation (Chincarini et al. 2010;
Bernardini et al. 2011), where $t_{\rm p}=t_{\rm p}/(1+z)$ is the cosmological rest-frame peak time of
the flares. On the other hand, Figure \ref{fig:correlation}(f) shows the $T_{\rm d}-t_{\rm p}$ diagram,
which indicates that the larger the peak time, the longer the duration of the X-ray flare.

\section{Possible theoretical models for interpreting the X-ray variability}
Based on above discussion, the X-ray variability of GRB 170817A seems shared the similar properties
with the X-ray flares observed in GRB afterglow. If this is the case, this X-ray variability can give a
strongly constraint with remnants of binary neutron star mergers. In this section, we present three
possible remnants (e.g., a stable magnetar, a long-lived supra-massive NS, and black hole) of this
double NSs mergers, and discuss several theoretical models to interpret this X-ray variability with
different remnants.

\subsection{A stable magnetar as central engine?}
One possible remnant of such neutron star mergers is a stable magnetar if its mass is below the
maximum mass of a non-rotating NS or the equation of state is stiff enough (Giacomazzo \& Perna
2013; Gao et al. 2016). Within this scenario, the most important issue is how to produce the X-ray
variability at late times. In a previous study, several models were proposed to interpret the X-ray
flare in short GRBs, e.g., differential rotation of the NS (Dai et al. 2006) or fallback accretion
onto a new-born magnetar (Gibson et al. 2018). In this section, we will elaborate on these two
models that may produce the X-ray variability in GRB 170817A.

\subsubsection{Differential rotation of NS}
A differentially rotating massive neutron star existing as the post-merger object after a double
neutron star merger is thought to be correct because of the existence of millisecond pulsars and
because it is seen in relativistic numerical simulations (Nice et al. 2005; Shibata et al. 2005).
Moreover, Dai et al. (2006) invoked this model to interpret the X-ray flare in short GRBs. Following
the method in Dai et al. 2006, by considering a two-component model (the core and shell components),
whose angular velocities are $\Omega_c$ and $\Omega_s$. Thus, the differential angular velocity is
$\bigtriangleup \Omega=\Omega_c-\Omega_s$. The relationship between radial magnetic field ($B_r$) and
the toroidal field ($B_{\phi}$) can be written as
\begin{eqnarray}
\frac{dB_{\phi}}{dt}=(\bigtriangleup\Omega)B_r,
\label{Magneticfield}
\end{eqnarray}
and the solution to $\bigtriangleup\Omega$ can be expressed as
\begin{eqnarray}
\bigtriangleup\Omega=A_0\Omega_{s,0}\cos(t/t_0),
\label{Omega}
\end{eqnarray}
where $\Omega_{s,0}$ is initial shell angular velocity, $A_0$ is the ratio between the initial
differential angular velocity and the shell angular velocity, and $t_0$ is a constant that depends on
the NS equation of state and the radial magnetic field. The toroidal field will float up toward the
stellar surface with time until it reaches a critical value ($\sim 10^{17}$ G; Kluzniak \& Ruderman
1998). At that time, the buoyancy force is enough to balance the interior antibuoyancy of the NS and
can be estimated as
\begin{eqnarray}
t_{\phi}=\frac{B_{\phi}}{B_rA_0\Omega_{s,0}}\sim 4.8\times
10^4(\eta/0.3)B^{-1}_{s,8}A^{-1}_{0}P_{0,-3}\rm~s,
\label{floattime}
\end{eqnarray}
where $B_{s,8}$ is the surface dipole field strength in units of $10^8$ G, $\eta$ is the ratio between
the surface dipole field strength and the radial field strength, and $P_{0,-3}$ is the initial spin
period of the shell component in units of $10^{-3}$ seconds\footnote{The convention $Q=10^{x}Q_{x}$ is
adopted in cgs units for all other parameters throughout the paper.}.

Observationally, the occurrence time of the X-ray variability in GRB 170817A is $\sim137$ days,
which is approximately equal to $t_{\phi}$. For given typical values of the model parameters (i.e.,
$A_0\sim1$, $P_0\sim 1$ ms, and $\eta\sim0.3$; Dai \& Lu 1998a; Shibata et al. 2005), the surface
magnetic field of a NS can be estimated as $B_s\sim4.07\times10^5$ G. Even for larger $P_0\sim 10$
ms, the surface magnetic field is increased to only $B_s\sim4.07\times10^6$ G. This field strength
is too low to be consistent with the surface magnetic fields of normal pulsars (Goldreich \&
Reisenegger 1992). Thus, there is difficulty in interpreting the X-ray variability in GRB 170817A
at late times via differential rotation of the remnant neutron star.

\subsubsection{Fallback accretion onto a new-born magnetar}
A highly magnetized neutron star surrounded by a hyper-accreting disk resulting from the fallback
of matter could occur during the merger of two neutron stars (Dai \& Lu 1998a,b; Zhang \& Dai 2008;
2009). Gibson et al. (2018) proposed that the flares in GRB afterglows can be powered by in-falling
matter accreted onto the magnetar surface when the in-falling material from the accretion disk
rotates faster than the magnetosphere, or when the magnetosphere's centrifugal drag can not
overcome gravity (Piro \& Ott 2011; Bernardini et al. 2013; Gompertz et al. 2014; Gibson et al.
2017). We assume that the X-ray variability in GRB 170817A is produced by the fallback accretion
onto a new-born magnetar. Then, we make a rough estimate of the accretion mass and surface dipole
field strength of the magnetar.

For the accretion of a disc or torus onto the newly formed magnetar, two radii are discussed (Piro \&
Ott 2011; Gompertz et al. 2014; Gibson et al. 2017). One is the Alfv\'en radius
($r_m=B_{\ast}^{4/7}R_{\ast}^{12/7}(GM_{\ast})^{-1/7}\dot{M}^{-2/7}$) at which the dynamics of the gas
within the disc become strongly influenced by the dipole field, where $\dot{M}$ is accretion rate, and
$R_{\ast}$, $M_{\ast}$,and $B_{\ast}$ are the radius, mass, and magnetic field of magnetar,
respectively. The other is the co-rotation radius ($r_c\sim
\frac{2}{3}\pi^{-2/3}G^{1/3}M_{\ast}^{1/3}P^{2/3}$) at which material orbits at the same rate as the
magnetar surface, where $G$ is gravitational constant and $P$ is the period. When $r_c>r_m$, the
accretion disc is rotating faster than the magnetic field and magnetic torques slow the in-falling
material, allowing it to accrete. However, when $r_c<r_m$, the magnetic field is rotating faster than
the disc, allowing particles to be accelerated to super-Keplerian velocities and be ejected from the
system.

Due to the strong surface magnetic field of the magnetar, the magnetic pressure at a given radius $r$
prevents material from falling in
\begin{eqnarray}
P_{\rm mag}=\frac{2\pi^3B_{\ast}^2R_{\ast}^6}{3c^4P^{4}r^2}.
\label{Pmag}
\end{eqnarray}
Material falling in from the accretion disc also exerts an inward ram pressure, opposing that of
$P_{\rm mag}$,
\begin{eqnarray}
P_{\rm ram}=\frac{\dot{M}}{8\pi}(\frac{2GM_{\ast}}{r^5})^{1/2}.
\label{Pram}
\end{eqnarray}
Now, we consider the special situation where the magnetic pressure and ram pressure are in equilibrium
at $r=r_m$, namely, $P_{\rm mag}=P_{\rm ram}$. On the other hand, $r_m$ should be equal to $r_c$ when
those two pressures are in equilibrium. Therefore one can derive the relationship among $B_{\ast}$,
$P$, and $\dot{M}$ to be
\begin{eqnarray}
B_{\ast}=3.16\times 10^{10} (M_{\ast})^{5/6}R_{\ast,6}^{-3}P_{-3}^{7/6}\dot{M}_{-4}^{1/2} \rm~G.
\label{PBM}
\end{eqnarray}
From the observational point of view, the total energy of X-ray variability is $E_{\rm X, iso}\sim
3.13\times10^{45}\rm ~erg$. This requires a total fall-back mass as low as $\sim 1.7\times
10^{-7}\rm~M_{\odot}$ for given efficiencies $\eta_1=0.1$ and  $\eta_2=0.1$, where $\eta_1$ and
$\eta_2$ are the efficiencies with which fall-back mass is accreted onto the NS surface and gets
converted into X-ray emission, respectively. By adopting $P=3$ ms, the GM1 EOS with
$M_{\ast}=2.37\rm~M_{\odot}$ and $R_{\ast}=12.05\rm~km$ (Lasky et al. 2014), one can estimate
$B_{\ast}\sim 1.04\times 10^{10}\rm~ G$. This value is much smaller than the value estimated in
short GRBs with a magnetar central engine (L\"{u} et al. 2015).

\subsection{Central engine with a long-lived supra-massive NS?}
Another possible merger remnant of such binary systems is a supra-massive NS when the neutron star
EOS is stiff enough (Zhang 2013; Gao et al. 2016). Within this scenario, one needs to test whether
or not the X-ray variability in GRB 170817A can be produced by a supra-massive NS via the expelled
magnetosphere following the collapse. From the observational point of view, the jet opening angle
of GRB 170817A can be estimated with $\theta_j\sim 4^{\circ}$ based on the properties of the
multi-wavelength afterglow even if no jet break feature is observed in afterglow emission at a
later time (Troja et al. 2018). If we consider the beaming corrected of the GRB outflow, the lower
limit energy of X-ray variability can be estimated as,
\begin{eqnarray}
E_{\rm X}=E_{\rm X, iso}f_{\rm b}\simeq2.2\times10^{44}\rm~erg ,
\label{Ex}
\end{eqnarray}
where $f_b$ is beaming factor of the outflow $f_b\simeq(1/2)\theta^2_j=0.07$. Moreover, the upper
limit of X-ray flare in GRB 170817A is the isotropic energy, which is $E_{\rm X}\simeq
3.13\times10^{45}\rm~erg$.

Within this collapsing supra-massive NS framework, the magnetic field energy after the magnetosphere is
expelled following the collapse of the NS (i.e. the energy budget of magnetosphere) can be estimated as
(Zhang 2014),
\begin{eqnarray}
E_{\rm B} &\approx& \int^{R_{\rm LC}}_{R}4\pi r^2
\frac{B^2_p}{8\pi}\left(\frac{r}{R}\right)^{-6} dr \nonumber \\& \approx &
(1/6)B^2_p R^3 \approx (1.7\times 10^{45})\, B^2_{p, 14} R^3_{6}~{\rm erg},
\label{EB}
\end{eqnarray}
where $R_{\rm LC}\gg R$ is the light-cylinder radius, and $B_p$ is the surface polar cap magnetic
field. It is assumed that the X-ray variability of GRB 170817A is produced by the ejection of the
magnetosphere from a supra-massive NS collapsing into a black hole. This requires that the total
amount of magnetospheric energy should be larger than the energy of X-ray variability, e.g.,
$E_{\rm B}\geq E_{\rm X}$. If this is the case, the range of the surface polar cap magnetic field
$B_p$ should be in $[3.6\times10^{13}-1.35\times10^{14}]$ G by adopting the GM1 EOS with $M_{\rm
TOV}=2.37M_{\odot}$ and $R = 12.05$ km (Lasky et al. 2014; L\"{u} et al. 2015). This value seems to
be consistent with that estimated by the late time multi-band EM observations (Ai et al. 2018; Yu
et al. 2018) when the ellipticity $\epsilon$ reaches $\sim10^{-3}$ (Lasky \& Glampedakis 2016).
Alternatively, Piro et al. (2019) proposed that this X-ray variability can be produced if the
toroidal magnetic field component ($B_{\rm t}\geq 10^{14}$ G) is much stronger than the poloidal
component in the NS.

\subsection{Black hole central engine?}
The last possible outcome of such neutron star mergers is a black hole if the nascent NS mass is much
greater than the maximum non-rotating mass (Rosswog et al. 2003; Rezzolla et al. 2011; Ravi \& Lasky
2014; Gao et al. 2016). In this scenario, the late X-ray variability in GRB 170817A may be produced by
fall-back accretion onto the black hole (Wu et al. 2013; Chen et al. 2017). Following the method in Wu
et al. (2013) and Chen et al. (2017), one can estimate the radius, mass, as well as the fall-back
accretion rate. The spin energy of the BH can be tapped by a magnetic field connecting to the outer
world through the Blandford \& Znajek mechanism (Blandford \& Znajek 1977, hereafter BZ); the jet
produced in such a scenario will be dominated by the BZ power at late times (Zhang \& Yan 2011; Lei \&
Zhang 2011; Lei et al. 2013; Liu et al. 2015; Chen et al. 2017).

The minimum radius around which matter starts to fall back, the accretion rate, and the total accreted
mass can be estimated as
\begin{eqnarray}
R_{\rm fb}\simeq 2.91\times10^{13}(M_{\bullet}/3M_{\odot})^{1/3}(t_{\rm fb}/100\rm~day)^{2/3}\rm~cm,
\label{fallback}
\end{eqnarray}
\begin{eqnarray}
\dot{M}_{p}\simeq 1.1\times10^{-15}L_{\rm X, iso,39}a^{-2}_{\bullet}X^{-1}(a_{\bullet})\eta^{-1}_{-2}f_{\rm
b,-2}\rm~M_{\odot}/s,
\label{accration}
\end{eqnarray}
\begin{eqnarray}
M_{\rm fb} \simeq \int^{t_p}_{t_0}\dot{M}dt\simeq 2\dot{M}_{p}(t_p-t_0)/3,
\label{mass}
\end{eqnarray}
where $M_{\bullet}$ is the BH mass, $t_{\rm fb}$ is the fall-back time scale which is approximately
equal to the free-fall time, $a_{\bullet}$ is the spin parameter of BH, $X$ is as function of
$a_{\bullet}$ with $X(0.1)=0.17$, $X(0.5)=0.20$ and $X(0.9)=0.42$, and $\eta$ is the efficiency of
converting BZ power to X-ray radiation. For the late X-ray variability in GRB 170817A, we assume
that the BH mass and the efficiency are $3\rm~M_{\odot}$ and $0.01$, respectively. By adopting
$t_{\rm fb}\simeq T_{s}\simeq t_0=137$ days, $L_{\rm X, iso}\sim1.54\times10^{39}\rm~erg~s^{-1}$,
and $f_b\sim0.07$, as well as $a_{\bullet}=0.1$, $0.5$ and $0.9$, one can estimate $R_{\rm fb}\sim
3.59\times10^{13}\rm~cm$, and the total fall-back mass $M_{\rm fb}\sim 8.37 \times
10^{-6}\rm~M_{\odot}$, $2.85 \times 10^{-7}\rm~M_{\odot}$, and $4.18 \times 10^{-8}\rm~M_{\odot}$.

In general, the starting fall-back radius of black hole central engine with mass of $3M_{\odot}$ is
about $\sim 10^{10}$ cm, which is about the radius of helium star (Wu et a. 2013). But it also
dependent on the parameters of black hole. In our estimation, the minimum radius require a much
larger radius $\sim 10^{13}$ cm. In other words, the accreted time is too long to be consistent
with the required of black hole. On the other hand, as shown in Piro et al. (2019), if we assume
that the accretion starts after the merger for given typical ejecta masses of NS mergers (e.g.,
$M_{\rm ejc}\sim 0.01\rm~M_{\odot}$), the X-ray luminosity at $t\sim 160$ days is much higher than
the observed X-ray luminosity ($\sim 3\times 10^{39}\rm~erg~s^{-1}$) if we adopt a fall-back decay
index of $-5/3$ (Rosswog 2007). This is also incompatible with observations of the late X-ray
variability in GRB 170817A.

\section{Conclusions}
The remnants of double NS mergers remain unknown, as it depends on the total mass of the system and the
poorly known NS equation of state. However, the gravitational wave radiation and the multi-wavelength
electromagnetic counterparts of the system, especially the late electromagnetic radiation, can provide
some constraints on the outcome for that merger event (L\"{u} et al. 2018). Piro et al. (2019) claimed
that a low-significance X-ray variability in GRB 170817A was observed by {\em Chandra} and {\em
XMM-Newton} at 155 days, and that variability is analogous to the one observed in GRB afterglows. In
this work, by estimating the parameters of several possible merger products and their mechanisms for
producing the observed X-ray variability in GRB 170817A, we are able to reach several interesting
conclusions.
\begin{itemize}
 \item The X-ray variability in GRB 170817A seems to share similar statistical correlations with
     X-ray flares in GRB afterglows, which we found by systematically comparing the properties
     between them.
 \item If the central engine of GW170817/GRB 170817A is a stable magnetar, and we invoke the
     mechanisms of differential rotation or fall-back accretion onto the NS to produce the later
     X-ray variability in GRB 170817A, then it is required that the NS has an extremely low surface
     magnetic field. This seems to be inconsistent with current observations.
 \item If the central engine of this event is a black hole, the fall-back accretion onto the black
     hole requires an extremely large envelope and long accretion timescale, both of which are not
     convincing.
 \item Invoking a central engine with a long-lived supra-massive NS seems to be consistent with
     observations. The later X-ray variability may be produced by a magnetosphere which is expelled
     following the collapse of the NS with the minimum $B_p\sim5.25\times10^{13}$ G.
\end{itemize}

Our analysis also poses a curious question: how can a collapsing supra-massive NS produce a flare with
a duration longer than tens of days? The answer may lie with a model in which the ejected magnetosphere
interacts with a blast wave, but such a calculation would be difficult. We expect that future numerical
simulations may provide the solution to this question.

Upon received the referee comments of this paper, we were drawn attention to Lin et al. (2019), who
performed an independent theoretical interpretation on the later X-ray flare of GRB 170817A. They
proposed that the X-ray flare of GRB 170817A can be interpreted with a slim accretion disc around a
post-merger compact object.

\section{Acknowledgements}
We acknowledge the use of the public data from the {\em Swift} data archive and the UK {\em Swift}
Science Data Center. We thank the referee for helpful suggestions that improved this paper. We also
thank Bing Zhang for helpful discussion. This work is supported by the National Natural Science
Foundation of China (grant Nos.11603006, 11533003, 11851304, and 11773010), Guangxi Science
Foundation (grant Nos. 2017GXNSFFA198008, 2016GXNSFCB380005, 2017AD22006, and 2018GXNSFGA281007).
The One-Hundred-Talents Program of Guangxi colleges, Bagui Young Scholars Program (LHJ), the high
level innovation team and outstanding scholar program in Guangxi colleges, and special funding for
Guangxi distinguished professors (Bagui Yingcai \& Bagui Xuezhe).


\end{document}